\documentclass[sigconf, authorversion]{acmart}

\AtBeginDocument{%
  \providecommand\BibTeX{{%
    \normalfont B\kern-0.5em{\scshape i\kern-0.25em b}\kern-0.8em\TeX}}}

\copyrightyear{2023}
\acmYear{2023}
\setcopyright{acmlicensed}\acmConference[e-Energy '23]{The 14th ACM International Conference on Future Energy Systems}{June 20--23, 2023}{Orlando, FL, USA}
\acmBooktitle{The 14th ACM International Conference on Future Energy Systems (e-Energy '23), June 20--23, 2023, Orlando, FL, USA}
\acmPrice{15.00}
\acmDOI{10.1145/3575813.3595198}
\acmISBN{979-8-4007-0032-3/23/06}

\usepackage{colortbl}
\definecolor{lGray}{gray}{0.9}

\theoremstyle{definition}
\newtheorem{definition}{Definition}[section]

\usepackage{caption}
\usepackage{subcaption}
\usepackage[utf8]{inputenc}
\usepackage{pgfplots}
\DeclareUnicodeCharacter{2212}{−}
\usepgfplotslibrary{groupplots,dateplot}
\usetikzlibrary{patterns,shapes.arrows}
\pgfplotsset{compat=newest}
\usepackage{hhline}
\usepackage{multirow}
\usepackage{enumitem}

\expandafter\def\expandafter\UrlBreaks\expandafter{\UrlBreaks
    \do\a\do\b\do\c\do\d\do\e\do\f\do\g\do\h\do\i\do\j%
    \do\k\do\l\do\m\do\n\do\o\do\p\do\q\do\r\do\s\do\t%
    \do\u\do\v\do\w\do\x\do\y\do\z\do\A\do\B\do\C\do\D%
    \do\E\do\F\do\G\do\H\do\I\do\J\do\K\do\L\do\M\do\N%
    \do\O\do\P\do\Q\do\R\do\S\do\T\do\U\do\V\do\W\do\X%
    \do\Y\do\Z\do\/\do-}

\begin{document}

\title{Appliance Detection Using Very Low-Frequency Smart Meter Time Series}

\author{Adrien Petralia}
\orcid{0000-0003-2987-9111}
\affiliation{%
  \institution{\small{EDF - Université Paris Cité}}
  \city{Paris}
  \country{France}
}
\email{adrien.petralia@gmail.com}

\author{Philippe Charpentier}
\orcid{0000-0002-3039-2485}
\affiliation{%
  \institution{\small{EDF}}
  \city{Palaiseau}
  \country{France}
}
\email{philippe.charpentier@edf.fr}

\author{Paul Boniol}
\orcid{0000-0001-8516-0123}
\affiliation{%
  \institution{\small{Université Paris Cité}}
  \city{Paris}
  \country{France}
}
\email{boniol.paul@gmail.com}

\author{Themis Palpanas}
\orcid{0000-0002-8031-0265}
\affiliation{%
  \institution{\small{Université Paris Cité - IUF}}
  \city{Paris}
  \country{France}
}
\email{themis@mi.parisdescartes.fr}

\renewcommand{\shortauthors}{A. Petralia, et al.}

\begin{abstract}

    In recent years, smart meters have been widely adopted by electricity suppliers to improve the management of the smart grid system. 
    These meters usually collect energy consumption data at a very low frequency (every 30min), enabling utilities to bill customers more accurately. 
    To provide more personalized recommendations, the next step is to detect the appliances owned by customers, which is a challenging problem, due to the very-low meter reading frequency.
    Even though the appliance detection problem can be cast as a time series classification problem, with many such classifiers having been proposed in the literature, no study has applied and compared them on this specific problem.
    This paper presents an in-depth evaluation and comparison of state-of-the-art time series classifiers applied to detecting the presence/absence of diverse appliances in very low-frequency smart meter data. 
    We report results with five real datasets. 
    We first study the impact of the detection quality of 13 different appliances using 30min sampled data, and we subsequently propose an analysis of the possible detection performance gain by using a higher meter reading frequency. 
    The results indicate that the performance of current time series classifiers varies significantly. 
    Some of them, namely deep learning-based classifiers, provide promising results in terms of accuracy (especially for certain appliances), even using 30min sampled data, and are scalable to the large smart meter time series collections of energy consumption data currently available to electricity suppliers.
    This paper appeared in ACM e-Energy 2023.
\end{abstract}





\maketitle
\pagestyle{plain}

\begin{figure}
    \centering
    \includegraphics[width=1\linewidth]{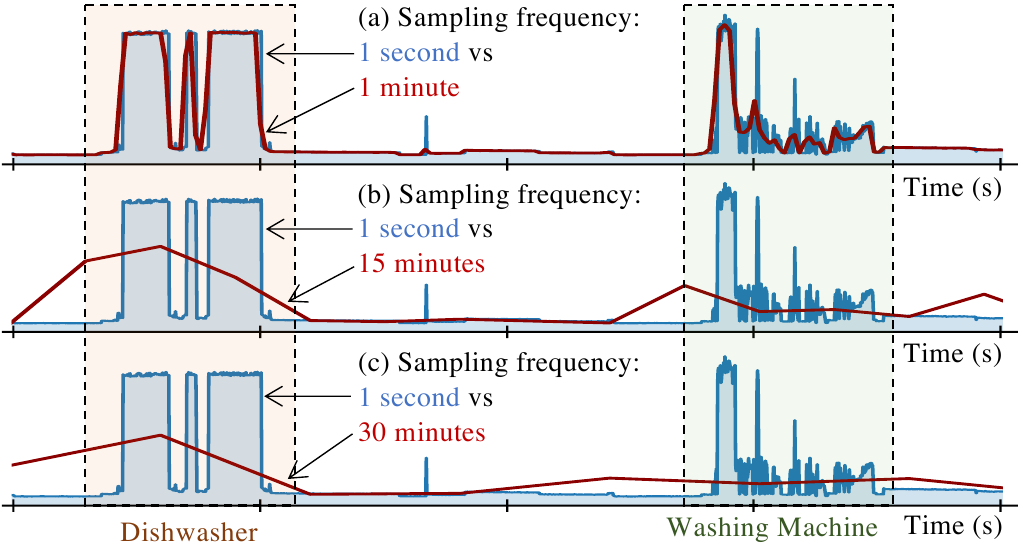}
    \vspace{-0.8cm}
    \caption{Comparisons of load curves containing a dishwasher and a washing machine at different sampling frequencies (1 second vs 1, 15, and 30min)}
    \label{fig:introfig}
\vspace{-0.6cm}
\end{figure}

\section{Introduction}
\label{sec:intro}

The energy sector is undergoing significant changes, primarily driven by the need for a more sustainable and secure energy supply. 
One way to better manage our consumption is to understand it better. 
In the last decade, electricity suppliers have installed millions of smart meters worldwide to improve their ability to manage the electrical grid~\cite{smart_meter_deployments_ue, smart_meter_deployments_us}.
These meters record detailed time-stamped data on electricity consumption, allowing both individual customers and businesses to better understand and rationalize their consumption~\cite{smartmeterinsmartgrid}.
These data are also valuable for suppliers, as they can help them anticipate energy demand more accurately. 
Overall, the widespread adoption of smart meters plays a crucial role in transitioning toward a more sustainable and efficient energy system.

We note that it has become essential for electricity suppliers to know which electrical appliances their customers own. 
This knowledge allows suppliers to better segment their customer base~\cite{ALLCOTT20111082}, and therefore to propose personalized offers and services that increase the customer satisfaction and retention.
Furthermore, they can help customers rationalize their electricity consumption, therefore contributing to the energy transition. 
One way to gather this information is by asking customers directly through a consumption questionnaire. 
However, this method can be a significant investment in terms of time and resources, which customers may not accept, and is also prone to errors. 
Therefore, electricity suppliers need to find more efficient and non-intrusive ways of gathering this information, such as using advanced data analytics techniques to detect the appliances directly through the collected smart meters data~\cite{192069}.

Appliance detection has become a significant area of research, with various techniques employed to detect the presence of devices~\cite{LIU2021283,ROSSIER2017691}. 
This problem is closely related to Non-Intrusive Load Monitoring (NILM), which aims to identify the power consumption, pattern, or on/off state activation of individual appliances using only the total consumption series~\cite{reviewnilm}. 
While detecting an appliance can be seen as a step in NILM-based methods~\cite{QU2023112749, ASLAN2022112087, 7498597, 6583496, 10.1145/3077839.3077845, https://doi.org/10.48550/arxiv.2209.03759, DBLP:conf/eenergy/LavironDHP21}, and diverse approaches have been proposed in the literature~\cite{QU2023112749, ASLAN2022112087, 7498597, 6583496, 10.1145/3077839.3077845, https://doi.org/10.48550/arxiv.2209.03759, DBLP:conf/eenergy/LavironDHP21}, they differ from our objective. 
Indeed, these studies essentially focus on detecting \emph{when} a specific appliance is "ON" rather than \emph{if} a household owns a specific appliance, and the presence of a specific appliance is in several cases already known before applying these approaches. 
Moreover, the majority of the NILM studies rely on data sampled at $\ge$1Hz, and consequently use signature-based methods~\cite{LIU2021283,ROSSIER2017691} that require either knowledge about how each appliance operates, or training on their individual power consumption.
Nonetheless, most existing smart meter installations record consumption at a very low sampling frequency: once every 10 to 60 minutes (in some cases at an even lower frequency). 
This results in signals where the unique appliance pattern information has been smoothed-out, or lost. 
Figure~\ref{fig:introfig} illustrates this loss of information. 
We observe that the dishwasher (shown on the left) and washing machine (shown on the right) signatures become increasingly hard to distinguish from one another as the sampling frequency drops.
Therefore, it becomes infeasible to accurately detect appliances using signature-based methods for the sampling frequencies actually used in practice.

In this paper, we 
propose a benchmark of diverse state-of-the-art classification methods for the problem of appliance detection in very low-frequency electrical consumption time series. 
We conduct our experimental evaluation on five real smart meter datasets using different time series classifiers. 
We first focus on detecting appliances in very low-sampled smart meters data (30min level), as it is nowadays one of the standard sampling rates adopted by electricity suppliers. 
We then provide an in-depth analysis of the increasing detection quality using higher frequency smart meter readings: 15min, 10min, and 1 min. 
To our knowledge, this is the first study to perform an exhaustive comparison of 11 state-of-the-art methods on five diverse real datasets with 13 different types of appliances, for multiple sampling frequencies.
The experimental evaluation demonstrates that current time series classifiers can accurately detect several appliances, even at the 30min resolution. 
Specifically, deep learning techniques are the most accurate and scalable when applied to large smart meter datasets.
Moreover, we demonstrate that setting the smart meter reading frequency to 1min can greatly enhance appliance detection using time series classifiers.

Our contributions are summarized as follows.
\begin{itemize}[noitemsep,topsep=0pt,parsep=0pt,partopsep=0pt,leftmargin=0.5cm]
    \item We describe a framework for comparing the performance of different time series classification methods for the appliance detection problem, and make this framework publicly available: \url{https://github.com/adrienpetralia/ApplianceDetectionBenchmark}
    \item We perform an extensive experimental evaluation using 5 diverse real datasets and 11 time series classifiers, including both traditional machine learning, as well as deep learning methods.
    \item We report the results of our comparison, which demonstrate that (i) current time series classifiers can only detect certain appliances at the 30min resolution; (ii) deep learning classifiers are the most accurate and scalable solution; and (iii) electricity suppliers should target a minimum smart meter reading frequency of 15min.
    \item The findings of this study can help electricity suppliers make informed decisions regarding the characteristics of future smart meter deployments. 
    Moreover, these findings point to interesting (and still challenging) open research directions in the context of electricity consumption time series analysis, and appliance detection in particular.
\end{itemize}


\section{Background and related work}
\label{sec:relatedwork}

\subsection{Smart Meter Data}

An electrical consumption load curve is defined as a univariate time series $\mathcal{X}=(\boldsymbol{x}_1, ..., \boldsymbol{x}_T)$ of ordered elements $\boldsymbol{x}_j \in \mathbb{R}_{+}^{1}$ following $(i_1,...,i_T)$ time consumption indexes (i.e., timestamps).
The sampling frequency is defined as the time difference between two records index $\Delta_t \coloneqq i_{j} - i_{j-1}$. Each element $\boldsymbol{x}_j$, usually given in Watt, indicates either the actual power at time $i_{j}$ or the average electric power called during the interval time $\Delta_t$.
The value can also be given in Watt-hour. 
In the literature, the definition of high and low-frequency smart meters data can differ~\cite{en14092390}. 
In this study, we refer to \emph{high-frequency} data sampled at less than 1 second and \emph{low-frequency} data sampled between 1 second and 1min. 
Data sampled above 1min refers to \emph{very low-frequency} smart meter data.

\noindent{\textbf{[Individual appliance load curve]}} By monitoring electric devices with individual meters, we can obtain the consumption load curve of each individual appliance in a household. 
However, instrumenting every appliance in the house is prohibitively expensive. 

\noindent{\textbf{[Aggregate load curve]}} The main consumption power of a house is usually recorded by a smart meter device located on the electrical meter of the household. 
This aggregate signal is the addition of the power consumption of all individual appliances in the household.

\subsection{Non-Intrusive Load Monitoring (NILM) and Appliance Detection}

Non-Intrusive Load Monitoring (NILM)~\cite{192069}, also called load disaggregation, relies on identifying the individual power consumption, pattern, or on/off state activation of individual appliances using only the total aggregated load curve~\cite{reviewnilm}.
NILM was initially approached as a problem involving linear combinations, with algorithms aiming to estimate the proportion of total power consumption used by distinct active appliances at each time step~\cite{reviewnilm}. 
Early research on this topic employed combinatorial optimization techniques~\cite{reviewnilm}. 
Later, Hidden Markov Models became the dominant approach, and in the last few years, deep learning models have been the reference to perform disaggregation~\cite{Kelly_2015, reviewnilm, en14092390, bert4nilm, electricitynilm}. 
Furthermore, NILM approaches can be divided into supervised and unsupervised learning, depending on whether they usee labeled data for training the models. 
Supervised learning involves classifying detected events (appliances being switched on or off) by matching extracted features~\cite{7498597, 6583496, LIU2021283,DBLP:conf/eenergy/LavironDHP21}. In contrast, unsupervised NILM methods detect events 
by analyzing feature similarities, or correlations without using labeled data~\cite{7457610, 192069}.

Since device recognition can be seen as a step of NILM-based methods, different approaches exist in the literature to detect appliances in load curves using high or low-frequency smart meter data~\cite{QU2023112749, ASLAN2022112087, 7498597, 6583496, 10.1145/3077839.3077845, https://doi.org/10.48550/arxiv.2209.03759,DBLP:conf/eenergy/LavironDHP21}. 
However, numerous studies using pattern recognition at low frequency require knowledge about how each device operates. 
Few recent research studies~\cite{https://doi.org/10.48550/arxiv.2209.03759, 6583496, ASLAN2022112087,DBLP:conf/eenergy/LavironDHP21} used time series features, or deep learning representations, to detect events or appliance activation patterns. 
Despite the promising results demonstrated by these studies using modern machine learning approaches, we note that they are only applied to high-frequency data (i.e., data sampled at a minimum rate of 1 sample per second).


\subsubsection{Studies on Very Low-Frequency Data}

Most NILM studies use high-frequency smart-meter data (seconds level at maximum), and only very few studies have been conducted using very-low sampling rates~\cite{c6cf02df821e4ae29ec3ea7f9fd66406, ZHAO2020114949, granularityinfluence}. 
In~\cite{ZHAO2020114949}, the authors suggested three methods to estimate appliance consumption using hourly smart meter data. 
The first two methods are unsupervised and require knowledge about manufacturer appliance parameters.
The third method is a supervised deep learning approach that requires disaggregate appliance load curves for training. 
In~\cite{granularityinfluence}, the authors proposed a data privacy-oriented study to assess the impact of Smart Meters sampling frequency on the detection of certain electrical appliances.
They used an event detection approach as a feature extractor to train a classifier that identifies changes in power consumption. 
However, the experimental evaluation in this study was rather limited: the authors evaluated the classification performance on a single dataset, and used the same house for both training and testing.
Overall, the few NILM studies that used low-frequency data focus on estimating the consumed power of each appliance, rather than detecting which appliances are present in the households.

Few papers in the literature~\cite{Albert_2013, 9299491} try to tackle the problem of detecting the devices owned by a household using very low-frequency sampled data.
In~\cite{Albert_2013}, the authors used a Hidden Semi-Markov Model (HSMM) to extract appliance features from power consumption data. 
These features are then merged with external variables (such as temperature) and serve to train an AdaBoost classifier~\cite{adaboost} to detect the presence of different appliances.
In~\cite{9299491}, the authors proposed a framework that uses a deep learning approach on subsequences of a long consumption load curve to detect the appliances present in the household. 
A majority vote gives the final device prediction, based on the individual predictions made on every examined subsequence. 
The study compares their method to~\cite{Albert_2013}, but not to any of the current state-of-the-art time series classifiers. 
In addition, only one public dataset at one sampling rate was considered.

\subsection{Time Series Classification}

Time series classification (TSC) \cite{https://doi.org/10.48550/arxiv.1602.01711, 10.1007/s10618-019-00619-1} is an important analysis task across several domains. 
Many studies have suggested different approaches to solve the TSC problem, ranging from the computation of similarity measures between time series~\cite{1053964} to the identification of discriminant patterns~\cite{Hills2014}. 
In addition, benchmarks, such as the the UCR archive~\cite{UCR2018}, have been proposed, on which exaustive experimental studies have been conducted~\cite{https://doi.org/10.48550/arxiv.1602.01711}.  
We discuss in more detail the current state-of-the-art time series classifiers in Section~\ref{sec:proposedbenchmark}.

\section{Problem Definition and Benchmark}
\label{sec:proposedbenchmark}


\subsection{Problem Definition}

In this work, we treat the appliance detection problem as a supervised binary classification problem. 
We aim to identify the presence/absence of a specified appliance's activation signature in a smart meter data series, independently of the number of activations of this appliance.
The presence can be simply defined by the fact that the device is switched "ON" at least once.
Formally, we define the problem as follows:

\begin{definition}[Appliance Detection Problem]
Given an aggregate smart meter time series $\mathcal{X} \in \mathbb{R}^{T}$, an  appliance type $a$, we want to know if appliance $a$ is activated at least once in $\mathcal{X}$ (i.e., was in an "ON" state, regardless of the time and number of activations).
\end{definition}


\subsection{Overview of Time Series Classifiers}
\label{sec:sotaclassifiers}

\begin{figure}
    \centering
    \includegraphics[width=1\linewidth]{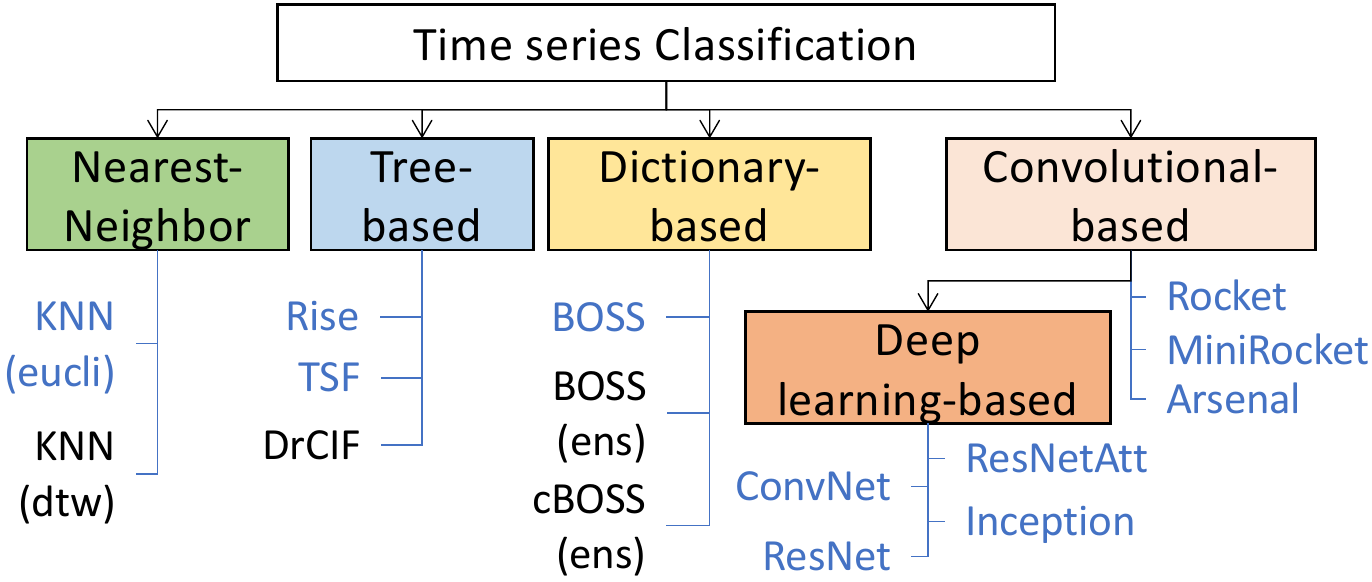}
    \vspace{-0.4cm}
    \caption{Taxonomy of classifier considered in our benchmark (in blue: classifier used in the experimental evaluation).}
    \label{fig:taxo}
\vspace{-0.2cm}
\end{figure}

We now provide an overview of the different approaches proposed in the literature to solve the TSC problem (refer to Figure~\ref{fig:taxo}). 
The objective is to compare the performance of these methods when applied to the appliance detection problem.
Each classifier takes as input for training the univariate consumption time series (i.e., 1D signal) along with the ground-truth labels.

\subsubsection{Nearest-Neighbor Classifier}

$K$-Nearest-Neighbor ($K$-NN) classifiers are the most simple and intuitive classifiers, based on the notion of time series similarity. 
Following a chosen distance measure, each new instance is classified by getting assigned the same label as the majority label of the $K$ closest samples in the training set ($K=1$ in our experiments, i.e., we use 1-NN classifiers). 
The most popular distance measure is Euclidean distance, which allows comparing two instances point to point. 
However, this distance does not consider the possible distortions on the temporal axis. 
Dynamic Time Warping (DTW)~\cite{1163055} is a distance measure to compute the similarity between two time series, where relevant patterns may evolve at different speeds. 
DTW suffers from a high computational cost, which makes it challenging to apply on large datasets. 

\subsubsection{Tree Based Classifier}

Tree-based classifiers, like Random Forest~\cite{randomforest}, have exhibited promising results in classification tasks. 

\noindent{\bf [Time Series Forest]}
TSF~\cite{https://doi.org/10.48550/arxiv.1302.2277} is a random forest-based classifier that uses as input features extracted from randomly sampled intervals of the raw data series. 
The algorithm first selects a number $r$ of intervals with a random start position and length; then, from each interval, three simple features are extracted: the mean, the standard deviation, and the slope. 
Finally, the $3r$ new features serve to train a classic random forest classifier.
The number of intervals is set by default to $\sqrt{T}$, where $T$ is the length of the input time series, and the number of estimators for the decision tree is set to $200$.

\noindent{\bf [Random Interval Spectral Ensemble]}
The RISE algorithm~\cite{hivecote1} is a random forest classifier based on spectral extraction features, rather than simple summary statistics for each interval. 
It computes the Fast Fourier Transform (FFT) and the Auto Correlation Function (ACF) for several randomly selected intervals. 
In contrast to TSF, the algorithm extracts only one interval from the raw series for each decision tree (set to $500$ in our experiments), and the first tree is built using the features extracted from the entire series.

\noindent{\bf [DrCIF]}
The Diverse Representation Canonical Interval Forest Classifier (DrCIF) algorithm~\cite{drCIF_clf} is an extension of the Canonical Interval Forest (CIF) classifier~\cite{cif}, which itself uses the Canonical Time Series Characteristics (Catch22)~\cite{catch22}. 
Unlike the two previous tree-based methods, this algorithm is an interval-based time series classifier that looks for discriminative subseries before building the decision trees.
As for TSF, the number of estimators is set to $200$ in our experiments.

\subsubsection{Dictionary Based Classifier}

Dictionary-based approaches, also called \emph{bag-of-words} approaches, transform a time series into a sequence of symbols (letters usually) according to a chosen discretization technique. 
Using a sliding window of a specific size $l$, it is then possible to count the number of repeated patterns (i.e., symbolic words) to perform classification regarding the repetition frequency of similar patterns.

\noindent{\bf [BOSS]}
The Bag Of SFA Symbol (BOSS)~\cite{bossclassifier} is a dictionary-based classifier that uses Symbolic-Fourier-Approximation (SFA)~\cite{sfa} as a discretization technique. 
It first extracts sub-sequences from the raw series using a predefined sliding window of length $l$. 
Then, each sub-series is discretized in a word of size $w$ of $\alpha$ symbols using SFA and the Multiple Coefficient Binning algorithms~\cite{bossclassifier} ($l=10$, $w=10$, and $\alpha=2$ in our experiments). 
This symbolic sentence (i.e., word arrangement) is then converted into a histogram by counting the frequency occurrence of each word. 
Finally, classification is performed using the histogram information. 

\noindent{\bf [BOSS and cBOSS Ensembles]}
The BOSS ensemble~\cite{bossclassifier} is a set of individual BOSS classifiers that use different discretization parameters $w$ and $l$. 
The parameter $l$ is defined as $l \in [ 10,T]$ ($T$ being the time series length), and values of $w \in \{16, 14, 12, 10, 8\}$.
The number of symbols, $\alpha$, is set to the default value of $4$. 
The algorithm keeps only individual BOSS classifiers that performed the best according to a validation test. 
The BOSS ensemble requires building and evaluating a large number of models, making it a time and memory-intensive classifier for large datasets. 
To address this complexity, a compact version (cBOSS) was introduced, that uses a restricted set of randomly chosen parameters for ensemble creation. 



\subsubsection{Deep Learning Based Classifier}

The interest in deep learning methods for time series classification has risen significantly in the past few years~\cite{https://doi.org/10.48550/arxiv.1611.06455,10.1007/s10618-019-00619-1}. 
These models have shown excellent performance, reaching the top of state-of-the-art. 

\noindent{\bf [ConvNet]}
A Convolutional Neural Network (CNN) \cite{DBLP:journals/corr/OSheaN15} is a type of deep learning neural network widely used in image recognition that is specially designed to extract patterns through data with a grid-like structure, such as images, or time series. 
A CNN uses convolution, where a filter is applied on a sliding window over the time series. 
The ConvNet architecture proposed in~\cite{https://doi.org/10.48550/arxiv.1611.06455} is composed of three stacked Convolutional blocks followed by global average pooling~\cite{https://doi.org/10.48550/arxiv.1312.4400}, and a Softmax activation function. 
Each Conv block comprises a convolutional layer followed by a batch normalization layer~\cite{https://doi.org/10.48550/arxiv.1502.03167}, and a ReLU activation layer. 
The three block used the following 1D kernel sizes $\{8, 5, 3\}$.

\noindent{\bf [ResNet]}
The Residual Network (ResNet) architecture~\cite{https://doi.org/10.48550/arxiv.1512.03385} was introduced to address the gradient vanishing problem encountered in large CNNs~\cite{simonyan2015a}. 
A ResNet is formed by stacking several blocks and connecting them together using residual connections (i.e., identity mapping). 
For time series classification, a ResNet architecture has been proposed in~\cite{https://doi.org/10.48550/arxiv.1611.06455}, and has demonstrated a strong classification accuracy~\cite{10.1145/3514221.3526183}. 
It is the same architecture as the previously described ConvNet model, with adding residual connection between each Convolutional block.

\noindent{\bf [ResNet with Attention Mechanism]}
In~\cite{9299491}, the authors proposed a an extension of the ResNet architecture to perform appliance detection. 
The model starts by extracting features using six convolution blocks with dilated convolution and residual connections, followed by two encoder/decoder modules that use a dot product attention mechanism. 
In this model, the dilated convolution (i.e., adding zeroes between the elements of the filter) aims to increase the receptive 
field of the kernels without increasing the number of parameters. 
After the feature extraction step, the classification step is performed using a multi-layer perceptron followed by a softmax activation function.

\noindent{\bf [InceptionTime]}
Inspired by inception-based networks in computer vision~\cite{inception2014}, an ensemble of five neural networks using Inception modules has been proposed for time series classification~\cite{Ismail_Fawaz_2020}.
The model consists of five identical networks using residual connections and convolutional layers. 
One network uses 3 Inception modules that replace the traditional residual blocks that we can find in a ResNet architecture. 
Each Inception modules consist of a concatenation of convolutional layers using different size of filters. Specifically, each module results in the following layers. 
In the case of multivariate time series, a 1D convolutional bottleneck layer is used to reduce the number of dimensions of the time series .
Then, the output is fed to three different 1D convolutional layers with different kernel sizes (10, 20, and 40) and one Max-Pooling layer with kernel size 3. 
The last step consists of concatenating the previous four layers along the channel dimension and applying a ReLu activation function to the output, followed by batch normalization. 
All the convolutional layers used in the module come with 32 filters and a stride parameter of 1.

\subsubsection{Random Convolutional Kernel Features Classifiers}

The authors of~\cite{DBLP:journals/corr/abs-1910-13051} proposed an approach based on convolution filters without learning any weights. 
Some variants of this model, based on the same principle, were later proposed in the literature.

\noindent{\bf[ROCKET]}
The RandOm Convolutional KErnel Transform (ROCKET) algorithm~\cite{DBLP:journals/corr/abs-1910-13051} uses 1D convolutional kernels to extract relevant features. 
Instead of learning proper filter parameters using a gradient descent algorithm to detect relevant patterns, the method generates a large set of $K$ kernels with random length, weights, bias, dilation, and padding. 
After applying them, the maximum and the proportion of positive values are extracted as new features for each time series, resulting in a $2K$ features for each instance. 
Classification is then performed on these features, using a simple ridge classifier. 
By default, ROCKET uses $10000$ random kernels.

\noindent{\bf[MiniRocket]}
MINImally RandOm Convolutional KErnel Transform (MiniRocket)~\cite{dempster_etal_2021} is a version of ROCKET 
that reduces the random sampling space of the filter parameters, and keeps only the proportion of positive values as a new feature for each kernel. 
These modifications lead to a lower execution time complexity while maintaining similar performances.

\noindent{\bf [Arsenal]}
Arsenal~\cite{hivecote2} is an ensemble of multiple ROCKET classifiers that uses a restricted number of kernels compared to the original model. 
This method was proposed to estimate the variance predicted by the classifier without changing the type of classifier.

\begin{table*}[tb]
\caption{Left side : datasets characteristics (number of time series, sampling frequency, time series length). Right side : selected appliance detection cases through the five datasets; for each case, the table summarizes the number of time series available ($\sharp$TS) and the imbalance degree of the test set for the case (IB Ratio). A slash indicate that no data are available for this case/dataset.}
\label{table:detaildataset}
\resizebox{\textwidth}{!}{
\begin{tabular}{c|c|c|c|c|c||c|c|cc|cc|cc|cc|cc}
    \toprule
    \multirow{3}{*}{\textbf{Datasets}} & \multirow{3}{*}{\textbf{Tot. TS}} & \multicolumn{4}{c||}{} & \multicolumn{2}{c|}{} & \multicolumn{10}{c}{\textbf{Datasets}} \\
     &  &  \multicolumn{4}{c||}{\textbf{TS Length}} & \multicolumn{2}{c|}{\textbf{Appliance case}} & \multicolumn{2}{c|}{REFIT} & \multicolumn{2}{c|}{UKDALE} & \multicolumn{2}{c|}{CER} & \multicolumn{2}{c|}{EDF 1} & \multicolumn{2}{c}{EDF 2} \\ 
     & & 1min & 10min & 15min & 30min & 
    \multicolumn{2}{c|}{} 
    & 
    $\sharp$TS & IB Ratio & 
    $\sharp$TS & IB Ratio & 
    $\sharp$TS & IB Ratio & 
    $\sharp$TS & IB Ratio & 
    $\sharp$TS & IB Ratio \\
    
    \midrule

    \multirow{2}{*}{REFIT} & \multirow{2}{*}{9091} & \multirow{2}{*}{1440} & \multirow{2}{*}{144} & \multirow{2}{*}{96} & \multirow{2}{*}{48} &
    \multirow{2}{*}{\rotatebox{90}{\scriptsize{Tech}}}
    & Desktop Computer & 5190 & 0.56 & \multicolumn{2}{c|}{$\diagup$} & 3286 & 0.47 & 1402 & 0.38 & 3740 & 0.62  \\
    & & & & & & & Television & 1134 & 0.92 & \multicolumn{2}{c|}{$\diagup$} & \multicolumn{2}{c|}{$\diagup$} & \multicolumn{2}{c|}{$\diagup$} & \multicolumn{2}{c}{$\diagup$} \\
    \hline

    \multirow{4}{*}{UKDALE} & \multirow{4}{*}{4767} & \multirow{4}{*}{1440} & \multirow{4}{*}{144} & \multirow{4}{*}{96} & \multirow{4}{*}{48} & 
    \multirow{4}{*}{\rotatebox{90}{\scriptsize{Kitchen}}}
    & Cooker & \multicolumn{2}{c|}{$\diagup$} & \multicolumn{2}{c|}{$\diagup$} & 1682 & 0.76 & \multicolumn{2}{c|}{$\diagup$} & \multicolumn{2}{c}{$\diagup$} \\
    & & & & & & & Kettle & 4790 & 0.72 & 1222 & 0.84 & \multicolumn{2}{c|}{$\diagup$} & \multicolumn{2}{c|}{$\diagup$} & \multicolumn{2}{c}{$\diagup$} \\
    & & & & & & & Microwave & 7434 & 0.55 & 1678 & 0.77 & \multicolumn{2}{c|}{$\diagup$} & 324 & 0.91 & \multicolumn{2}{c}{$\diagup$} \\
    & & & & & & & Electric Oven & \multicolumn{2}{c|}{$\diagup$} & \multicolumn{2}{c|}{$\diagup$} & \multicolumn{2}{c|}{$\diagup$} & 510 & 0.85 & 1152 & 0.91 \\
    \hline

    \multirow{3}{*}{CER} & \multirow{3}{*}{4225} & \multirow{3}{*}{$\diagup$} & \multirow{3}{*}{$\diagup$} & \multirow{3}{*}{$\diagup$} & \multirow{3}{*}{25728} & 
    \multirow{3}{*}{\rotatebox{90}{\scriptsize{Washer}}}
    & Dishwasher & 7798 & 0.44 & 2378 & 0.32 & 2350 & 0.66 & 224 & 0.93 & 2846 & 0.75 \\
    & & & & & & & Tumble Dryer & 3466 & 0.22 & \multicolumn{2}{c|}{$\diagup$} & 2214 & 0.68 & 1534 & 0.41 & 3470 & 0.42 \\
    & & & & & & & Washing Machine & 7422 & 0.54 & 2830 & 0.38 & \multicolumn{2}{c|}{$\diagup$} & \multicolumn{2}{c|}{$\diagup$} & \multicolumn{2}{c}{$\diagup$} \\
    \hline

    \multirow{3}{*}{EDF 1} & \multirow{3}{*}{2611} & \multirow{3}{*}{$\diagup$} & \multirow{3}{*}{$\diagup$} & \multirow{3}{*}{$\diagup$} & \multirow{3}{*}{17520} & 
    \multirow{3}{*}{\rotatebox{90}{\scriptsize{Heating}}}
    & Water Heater & \multicolumn{2}{c|}{$\diagup$} &  \multicolumn{2}{c|}{$\diagup$} & 3070 & 0.56 & 1336 & 0.66 & 548 & 0.86  \\
    & & & & & & & Electric Heater & \multicolumn{2}{c|}{$\diagup$} & \multicolumn{2}{c|}{$\diagup$} & 1348 & 0.19 & 1624 & 0.58 & 1538 & 0.56 \\
    & & & & & & & Convector/Heat Pump & \multicolumn{2}{c|}{$\diagup$} & \multicolumn{2}{c|}{$\diagup$} & \multicolumn{2}{c|}{$\diagup$} & 506 & 0.69 &  \multicolumn{2}{c}{$\diagup$} \\ \hline

    EDF 2 & 1553 & $\diagup$ & 26208 & 17472 & 8736 &
    \rotatebox{90}{\scriptsize{Other}} & Electric Vehicule & \multicolumn{2}{c}{$\diagup$} & \multicolumn{2}{c|}{$\diagup$} & \multicolumn{2}{c|}{$\diagup$} & 140 & 0.3 & \multicolumn{2}{c}{$\diagup$} \\
    \bottomrule
\end{tabular}
} 
\end{table*}

\subsubsection{Ensemble Models}

To reduce the variance in predictions, using a combination of models rather than a single one is a common technique. 
Ensemble models combining different approaches have been proposed to address the TSC problem. 
Several ensemble methods have been proposed in the literature, such as TS-CHIEF (Time Series Combination of Heterogeneous and Integrated Embedding Forest)~\cite{DBLP:journals/corr/abs-1906-10329} and HIVE-COTE (Hierarchical Vote Collective of Transformation-Based Ensembles)~\cite{hivecote2}. 
The first, is and ensemble of tree classifiers. The second is combining 4 different classifiers and use majority voting to provide the final prediction. 
However, these models suffer from a high execution time and cannot be applied to very long time series such as load curves.



\subsection{Energy Consumption Datasets}

Numerous energy consumption datasets exist in the literature~\cite{iedldataset}, and some of them have become references to conduct NILM studies~\cite{Kolter2011REDDA, ukdale, refitdataset}. 
However, these datasets typically provide aggregated and appliance-level load curves for only a few houses at a high-sampling frequency. Resampling them at a very low frequency leads to significant data reduction. 
In order to include a broader range of appliances and to align with existing literature, we include two NILM datasets in our experiments: UK-DALE~\cite{ukdale} and REFIT~\cite{refitdataset}.
We also include one public dataset providing 30min sampled aggregate load curves for a large number of households~\cite{CER_2012}. 
Moreover, we include two private datasets from EDF (the main french electricity supplier). 
In total, we consider five real diverse datasets in our experimental evaluation. 
These datasets are detailed below.


\subsubsection{NILM Datasets} UKDALE and REFIT are two well-known high-frequency Smart Meters datasets used in 
NILM studies~\cite{bert4nilm, electricitynilm}. 

\noindent{\textbf{[UK-DALE]}} The UK-DALE dataset~\cite{ukdale} contains data from 5 houses in the United Kingdom, and includes appliance-level load curves sampled every 6 seconds, as well as the whole-house aggregate data series sampled at 16kHz. 
Four houses were recorded for over a year and a half, while the 5th house was recorded for 655 days.

\noindent{\textbf{[REFIT]}} The REFIT project (Personalised Retrofit Decision Support Tools for UK Homes using Smart Home Technology)~\cite{refitdataset} ran between 2013 and 2015. 
During this period, 20 houses in the United Kingdom were recorded after being monitored with smart meters and multiple sensors. 
This dataset provides aggregate and individual appliance load curves at 8-second sampling intervals.

\subsubsection{CER Dataset}

The Commission for Energy Regulation of Ireland conducted a study to assess the performance of smart meters and their impact on consumer energy consumption~\cite{CER_2012}, recording the aggregate load curve consumption every 30min for over 5000 Irish homes and businesses. 
Pparticipants filled out a questionnaire on the household composition, the behavior of electricity consumption, and the type and number of appliances present in the home, or business. 
In this work, we use the residential sub-group of the study, i.e., 4225 households recorded from July 15, 2009, to January 1, 2011, for a total of 4225 
series, of length 25728 data points each.

\subsubsection{EDF Datasets}

To better understand its customers' base and electricity consumption behavior, \textit{Electricité De France (EDF)} conducts surveys on customer samples. 
These customers consent to EDF to use their data and analyze their consumption behaviors, and only the aggregate power consumption of the house is recorded. 
Similar to the CER study, customers fill out a questionnaire with information on which appliances are present in their households, and on their consumption habits. 
Two EDF datasets from two different studies were used in our experiments.

\noindent{\textbf{[EDF Dataset 1]}} The first one contains 2611 load curves at 30min sampling frequency of one year of electricity recording consumption. 
Data were collected between September 2019 and September 2021 from 1553 different clients. 
The dataset consists of 2611 time series of length 17520 from 1553 different sources.

\noindent{\textbf{[EDF Dataset 2]}} The second dataset contains 5354 load curves at a 10min sampling frequency, recorded over a period of six months. 
Data were collected between January 2012 and January 2015 from 1260 clients. 
The dataset consists of 5354 time series of length 26208 from 1260 different sources.

\section{Experimental Setup}
\label{sec:expsetup} 


All experiments are performed on a high-performance computing cluster. 
The source code is in Python 3.7, and for each classifier 
we use the default parameters provided by the authors in the original papers.
For non-deep-learning approaches, we use the sktime library~\cite{https://doi.org/10.48550/arxiv.1909.07872}. 
We perform each experiment on a server with 2 Intel Xeon Gold 6140 CPUs with 190 Go RAM. 
For deep-learning based models, we implement all the models using the 1.8.1 version of the PyTorch framework~\cite{https://doi.org/10.48550/arxiv.1912.01703}, and run experiments on a server with 2 NVidia V100 GPUs with 16Go RAM.

We consider all the classifiers presented in Section~\ref{sec:sotaclassifiers}. 
We run each method five-time using different random train/validation/test splits, and we report the average of these runs.
Note that the error bars shown in Figure~\ref{fig:ClfScore30min}, Figure~\ref{fig:ClfScoreall}, and Figure~\ref{fig:TSLenImpact}, correspond to the average variability of the classifiers through these five runs.
Additionally, we set a 10-hour time limit per job. 
Only models that finished a run (training + inference) are considered. 
We note that the ResNet with Attention model was not evaluated using UKDALE and REFIT data due to the residual block's dilation convolution being incompatible with the small size of the time series of these datasets. 

We make all code available online: \url{https://github.com/adrienpetralia/ApplianceDetectionBenchmark}

\begin{table*}[tb]
\caption{Results (average Macro F1-score for 5 runs) for the 11 classifiers (as well as the average score of all classifiers) evaluated through the appliance detection cases (best in bold and second best underlined). The "Appliance Average Score" row shows the average detection score for a specific device detection case if the appliance is available on multiple datasets. A slash indicates that the corresponding classifier failed to run on this case (time series length was not sufficiently large).}
\label{table:results30min}
{\footnotesize
\centering
\setlength\tabcolsep{2pt}
\begin{tabular}{cl||ccccccccccc|c}
\toprule
 {\small \textbf{Appliance}} & {\small \textbf{Dataset}}  &  {\small \textbf{Arsenal}} & {\small  \textbf{Minirocket}} & {\small \textbf{Rocket}} &  {\small \textbf{ConvNet}} &  {\small \textbf{ResNet}} &  {\small \textbf{ResNetAtt}} &  {\small \textbf{InceptionTime}} &   {\small \textbf{BOSS}} &  {\small \textbf{TSF}} &   {\small \textbf{Rise}} &  {\small \textbf{KNNeucli}}  & {\small \textbf{Avg. Score}} \\
\hline

\multirow{5}{*}{Desktop Computer} 
                & CER &    0.618 &       0.617 &   0.606 &    0.602 &   0.614 &       0.530 &      0.608 &  0.516 &             0.580 &  0.586 &     0.491 & 0.579\\
                & EDF 1 &    0.571 &       0.564 &   0.570 &    0.489 &   0.560 &       0.459 &      0.555 &  0.491 &             0.533 &  0.543 &     0.469 & 0.528 \\
                & EDF 2 &    0.603 &       0.576 &   0.582 &    0.579 &   0.620 &       0.514 &      0.601 &  0.519 &             0.570 &  0.592 &     0.520 & 0.571 \\
                & REFIT &    0.697 &       0.683 &   0.674 &    0.715 &   0.740 &         $\diagup$ &      0.623 &  0.542 &             0.525 &  0.600 &     0.548 & 0.635\\
\multicolumn{2}{c||}{\it{Appliance Average Score}} & \underline{0.622} &       0.610 &   0.608 &    0.596 &   \textbf{0.634} &  $\diagup$ &      0.597 &  0.517 & 0.552 &  0.580 &  0.507 & 0.578 \\
             \hline

Television  & REFIT & 0.656 &  0.647 & 0.645 &  0.695 & 0.699 &  $\diagup$ &  \underline{0.718} &  0.485 &   \textbf{0.737} &  0.664 & 0.513 & 0.646 \\
            \hline

Cooker & CER &    0.680 &       0.673 &   0.676 &    0.661 &   \underline{0.689} &       0.541 &      \textbf{0.710} &  0.526 &             0.566 &  0.584 &     0.440 & 0.613 \\
            \hline

\multirow{2}{*}{Kettle} & REFIT &    0.368 &       0.376 &   0.381 &    0.522 &   0.477 &         $\diagup$ &      0.415 &  0.536 &             0.359 &  0.428 &     0.421 & 0.428 \\
             & UKDALE &    0.540 &       0.502 &   0.522 &      0.428 	 &   0.432 &         $\diagup$ &      0.583 &  0.504 &             0.353 &  0.442 &     0.446 & 0.475 \\
             
\multicolumn{2}{c||}{\it{Appliance Average Score}} &    0.454 &       0.439 &   0.452 &    0.475 &   0.454 &         $\diagup$ &      \underline{0.499} &  \textbf{0.520} &             0.356 &  0.435 &     0.434 & 0.452 \\
            \hline
             
\multirow{3}{*}{Microwave} 
             & REFIT &    0.656 &       0.598 &   0.588 &    0.745 &   0.679 &         $\diagup$ &      0.673 &  0.563 &             0.540 &  0.717 &     0.529 & 0.629 \\
             & UKDALE &    0.446 &       0.498 &   0.460 &      0.532 &   0.526 &         $\diagup$ &      0.541 &  0.435 &             0.459 &  0.430 &     0.378 & 0.471  \\
             & EDF 1 &    0.480 &       0.471 &   0.475 &    0.534 &   0.510 &       0.409 &      0.474 &  0.454 &             0.400 &  0.429 &     0.457 & 0.463 \\
\multicolumn{2}{c||}{\it{Appliance Average Score}} &    0.527 &       0.522 &   0.508 &    \textbf{0.604} &   \underline{0.572} & $\diagup$ &      0.563 &  0.484 &             0.466 &  0.525 &     0.455 & 0.521 \\
            \hline
            
\multirow{2}{*}{Oven} 
             & EDF 1 &    0.513 &       0.498 &   0.499 &    0.512 &   0.512 &       0.472 &      0.523 &  0.506 &             0.429 &  0.497 &     0.437 & 0.491 \\
             & EDF 2 &    0.557 &       0.584 &   0.553 &    0.571 &   0.562 &       0.560 &      0.576 &  0.495 &             0.459 &  0.491 &     0.397 & 0.528 \\
\multicolumn{2}{c||}{\it{Appliance Average Score}} &    0.535 &       0.541 &   0.526 &    \underline{0.542} &   0.537 &       0.516 &      \textbf{0.550} &  0.500 &             0.444 &  0.494 & 0.417 & 0.509 \\
            \hline
             
\multirow{5}{*}{Dishwasher} 
             & REFIT &    0.650 &       0.599 &   0.619 &    0.580 &   0.605 &         $\diagup$ &      0.590 &  0.557 &             0.519 &  0.584 &     0.515 & 0.582 \\
             & UKDALE &    0.458 &       0.465 &   0.465 &      0.419 &   0.380 &         $\diagup$ &      0.384 &  0.399 &             0.429 &  0.554 &     0.525 & 0.448\\
             & CER &    0.699 &       0.720 &   0.700 &    0.730 &   0.728 &       0.594 &      0.737 &  0.586 &             0.609 &  0.648 &     0.488 & 0.658 \\
             & EDF 1 &    0.454 &       0.441 &   0.450 &    0.528 &   0.522 &       0.383 &      0.535 &  0.430 &             0.418 &  0.421 &     0.211 & 0.436\\
             & EDF 2 &    0.753 &       0.760 &   0.741 &    0.799 &   0.801 &       0.585 &      0.835 &  0.596 &             0.603 &  0.600 &     0.512 & 0.690\\
\multicolumn{2}{c||}{\it{Appliance Average Score}} &    0.603 &       0.597 &   0.595 &    \underline{0.611} &   0.607 &  $\diagup$ &      \textbf{0.616} &  0.514 &  0.516 &  0.561 &     0.450 & 0.563 \\
            \hline

\multirow{4}{*}{Tumble Dryer} 
             & REFIT &    0.493 &       0.503 &   0.502 &    0.468 &   0.448 &         $\diagup$ &      0.441 &  0.506 &             0.416 &  0.434 &     0.461 & 0.467 \\
             & CER &    0.634 &       0.641 &   0.628 &    0.606 &   0.612 &       0.550 &      0.623 &  0.549 &             0.578 &  0.602 &     0.474 & 0.591 \\
             & EDF 1 &    0.619 &       0.578 &   0.607 &    0.624 &   0.607 &       0.475 &      0.636 &  0.550 &             0.537 &  0.563 &     0.487 & 0.571 \\
             & EDF 2 &    0.733 &       0.714 &   0.714 &    0.757 &   0.769 &       0.475 &      0.769 &  0.560 &             0.593 &  0.681 &     0.493 & 0.660 \\
\multicolumn{2}{c||}{\it{Appliance Average Score}} &    \textbf{0.620} &       0.609 &   0.613 &    0.614 &   0.609 &       $\diagup$ &      \underline{0.617} &  0.541 &             0.531 &  0.570 &     0.479 & 0.572 \\
            \hline
            
\multirow{2}{*}{Washing Machine} & REFIT &    0.605 &       0.572 &   0.592 &    0.581 &   0.586 &         $\diagup$ &      0.614 &  0.520 &             0.562 &  0.557 &     0.529 & 0.572 \\
             & UKDALE &    0.475 &       0.505 &   0.478 &      0.535 &   0.530 &         $\diagup$ &      0.454 &  0.408 &             0.581 &  0.549 &     0.509 & 0.502 \\
\multicolumn{2}{c||}{\it{Appliance Average Score}} &    0.540 &       0.538 &   0.535 &    \underline{0.558} &   \underline{0.558} &         $\diagup$ &      0.534 &  0.464 &             \textbf{0.572} &  0.553 &     0.519 & 0.537 \\
            \hline
             
\multirow{4}{*}{Water Heater} & CER &    0.625 &       0.613 &   0.613 &    0.610 &   0.612 &       0.465 &      0.637 &  0.527 &             0.596 &  0.584 &     0.462 & 0.577\\
             & EDF 1 &    0.835 &       0.821 &   0.827 &    0.814 &   0.828 &       0.768 &      0.841 &  0.670 &             0.713 &  0.805 &     0.591 & 0.774 \\
             & EDF 2 &    0.733 &       0.685 &   0.724 &    0.731 &   0.685 &       0.591 &      0.759 &  0.658 &             0.580 &  0.666 &     0.617 & 0.675\\
\multicolumn{2}{c||}{\it{Appliance Average Score}} &    \underline{0.731} &       0.706 &   0.721 &    0.718 &   0.708 &       0.608 &      \textbf{0.746} &  0.618 &             0.630 &  0.685 &     0.557 & 0.675 \\
            \hline

\multirow{3}{*}{Heater} 
             & CER &    0.522 &       0.532 &   0.514 &    0.533 &   0.508 &       0.477 &      0.565 &  0.459 &             0.492 &  0.527 &     0.397 & 0.502 \\
             & EDF 1 &    0.784 &       0.783 &   0.789 &    0.777 &   0.778 &       0.713 &      0.800 &  0.643 &             0.758 &  0.777 &     0.638 & 0.749 \\
             & EDF 2 &    0.591 &       0.566 &   0.578 &    0.626 &   0.637 &       0.527 &      0.648 &  0.497 &             0.591 &  0.605 &     0.451 & 0.574 \\
\multicolumn{2}{c||}{\it{Appliance Average Score}} &    0.603 &       0.597 &   0.595 &    \textbf{0.659} &   0.607 &  0.572  & \underline{0.616} &  0.514 & 0.516 &  0.561 &     0.450 & 0.609 \\
\hline
            
Type of Heater & EDF 1 &    0.632 &       0.622 &   0.631 &    0.597 &   \underline{0.638} &  0.534 &      \textbf{0.651} &  0.539 &  0.556 &  0.625 & 0.467 & 0.590 \\
            \hline

Electric Vehicle & EDF 1 &    0.689 &       \textbf{0.730} &   0.670 &    0.681 &   0.699 &       0.553 &      0.720 &  0.541 &             0.456 &   \underline{0.725} &     0.556 & 0.638 \\
            
\hhline{==============}
\multicolumn{2}{c||}{\textbf{Classifiers Average Score}} & 0.601 &  0.593 &  0.592 &  0.609 &  \underline{0.610} &  $\diagup$  &  \textbf{0.617} &  0.521 & 0.531 &  0.574 & 0.474 & $\diagup$ \\
\multicolumn{2}{c||}{\textbf{Classifiers Average Rank}} & 3.773 &  4.697 &  4.758 &  4.303 &  \underline{3.697} &  $\diagup$  &  \textbf{2.864} &  7.939 & 7.924 &  6.197 & 8.848 & $\diagup$ \\
\bottomrule
\end{tabular}
} 
\end{table*}

\subsection{Data Preprocessing}

Since the datasets we employ in this study have been created using different sampling frequencies, we preprocess them for the experiments as explained below.
The left part of Table~\ref{table:detaildataset} summarizes, for each dataset, the number of time series and the corresponding length, according to each sampling frequency.

\subsubsection{NILM dataset preprocessing} 
The REFIT and UKDALE datasets provide appliance level and total consumption load curves for a small number of houses: 5 and 20, respectively. 
Moreover, the electrical appliances in the houses are likely the same. 
Inspired by the data processing step in NILM studies~\cite{bert4nilm, electricitynilm}, we preprocess the datasets by slicing the entire consumption curve of each household into smaller sub-sequences.

For each experiment, we first resample the data to a specified sampling rate and fill in with  linear interpolation the gaps of less than 1 hour. 
Then, we process the datasets by splitting each household's consumption load curve into smaller sub-sequences of one day, and by dropping those with missing values. 
The choice of the one day for the sub-sequence length provides an overall balance between positive (i.e., containing the device) and negative (i.e., not containing the device) samples.
In contrast, slicing the entire consumption curve in weeks leads to very few negative samples for most appliance cases. 
This is because the appliances in these datasets are devices that are very frequently used (on average, once every two or three days).
To assign the positive or negative label (i.e., appliance presence or not) to a sub-sequence, we use the corresponding disaggregated appliance load curve, allowing us to know if the appliance has been switched on at least once for a given day.

By preprocessing the UKDALE dataset, we noticed that the fourth house of the study could not be used for the experiments, since a single disaggregated load curve regrouped multiple appliances. 
Thus, we use only three houses for the training/validation set, whereas the one last house's sub-sequences are used for the test set.
With the REFIT data, we use two randomly selected houses for the test set, while the other houses available are used for the train set.

\subsubsection{CER and EDFs datasets preprocessing} 
The CER and EDFs datasets provide only the total aggregated load curve of each house. 
As a consequence, it is impossible to know if an appliance is activated or not for a given day. 
Therefore, we cannot slice the time series into smaller subsequences as for the NILM datasets, and we provide as inputs to the classifier the full-length load curves.
In addition, we process the load curves by linearly interpolating gaps of less than 1 hour and any time series with residual missing values are not retained.
The appliance presence label is assigned using the provided questionnaire associated with each dataset. 
Finally, we do a 70\%/10\%/20\% random split of the houses for the training, validation, and test sets, respectively.

\subsubsection{Appliance Detection Cases}

We select different cases of device detection through all the datasets, including small and big appliances. 
The right part of Table~\ref{table:detaildataset} summarizes the selected appliance detection cases for all datasets. 
The REFIT and UKDALE datasets include mostly small appliances because, in these studies, only plugged devices were recorded. 
On the other hand, the CER and EDFs datasets provide information about larger appliances, directly connected to the electric meters, such as Water Heaters, Heaters, and Electric Vehicles.

The selected cases aim to determine if a specific device is present in a time series using binary detection. 
However, the "Convector/Heat Pump" case involves classifying the types of electric heaters, such as distinguishing between convectors and heat pumps.

In order to ensure that the classifiers are not biased during training, we maintain an equal balance of time series labeled with positive and negative samples.
However, we note that the test set reflects the actual, imbalanced nature of the data, allowing us to evaluate the classifier's performance in a realistic scenario. 

\textbf{$\sharp$TS} is the number of labeled time series used for each case, in which each class are balanced. 
\textbf{IB Ratio} indicates the imbalance level of the corresponding test sets (i.e., the percentage of positive instances in the number of instances).

\subsection{Evaluation Measures}



\noindent{\bf[Accuracy]}
When detecting appliance presence/absence, several classification cases may be unbalanced. 
Indeed, most people own a television or a washing machine but do not have an electric heating system or a swimming pool. 
However, using a model that only predicts the majority class may appear to perform well in these cases when using the classification accuracy (i.e., the ratio of well classified instances versus the total number of instances). 
Precision, Recall, and the harmonic average of both, called the F1-Score, are well-known measures, 
defined as follows:
\begin{equation*}
    \text{F1-score} = \frac{2.P . R}{P + R} \text{, with: } P = \frac{TP}{TP+FP} \text{, } R = \frac{TP}{TP+FN}
\end{equation*}
with $TP = \text{True Positive}$, $TN = \text{True Negative}$, $FP = \text{False Positive}$ and $FN = \text{False Negative}$.
Nevertheless, precision (P), recall (R), and F1-score measures independently indicate the model's performance can be applied to one class only. 
In the case of a binary classification problem with data imbalance, these measures are typically applied only to the minority class. 
In our classification problem, the minority class varies depending on the specific device. 
Detecting an appliance (i.e., the positive class) could correspond either to the minority or the majority class. 
Thus, the F1-Score measure is not appropriate in our case. 
To account for this variability and provide an overall performance measure, we use the Macro F1-score to evaluate the performance of the classifiers. 
Formally, for $N$ class (in our case, $N=2$), the Macro F1-Score is defined as follows:
\begin{equation*}
    \text{Macro F1-score} = \frac{1}{N} \sum_{i=1}^{N} \text{F1-score}_i
\end{equation*}

\noindent{\bf[Time Performance]}
Considering the computation time of classifiers is crucial for evaluating their effectiveness in real-world scenarios. 
We measure the time performance of the classifiers, considering the total time required for both training and inference. 

\section{Results and Discussion}
\label{sec:resandanalysis}

This section presents the results of our experimental evaluation.
First, we normalize the different datasets to the same sampling frequency, i.e., 30min,  to obtain overall results on all the cases.
Then, we perform an experimental evaluation of the influence of sampling frequency on the detection quality of the classifiers.
We also analyze the data size impact on the detection quality.
Finally, we provide a discussion of the overall results.

\begin{figure}
    \centering
    \includegraphics[width=1\linewidth]{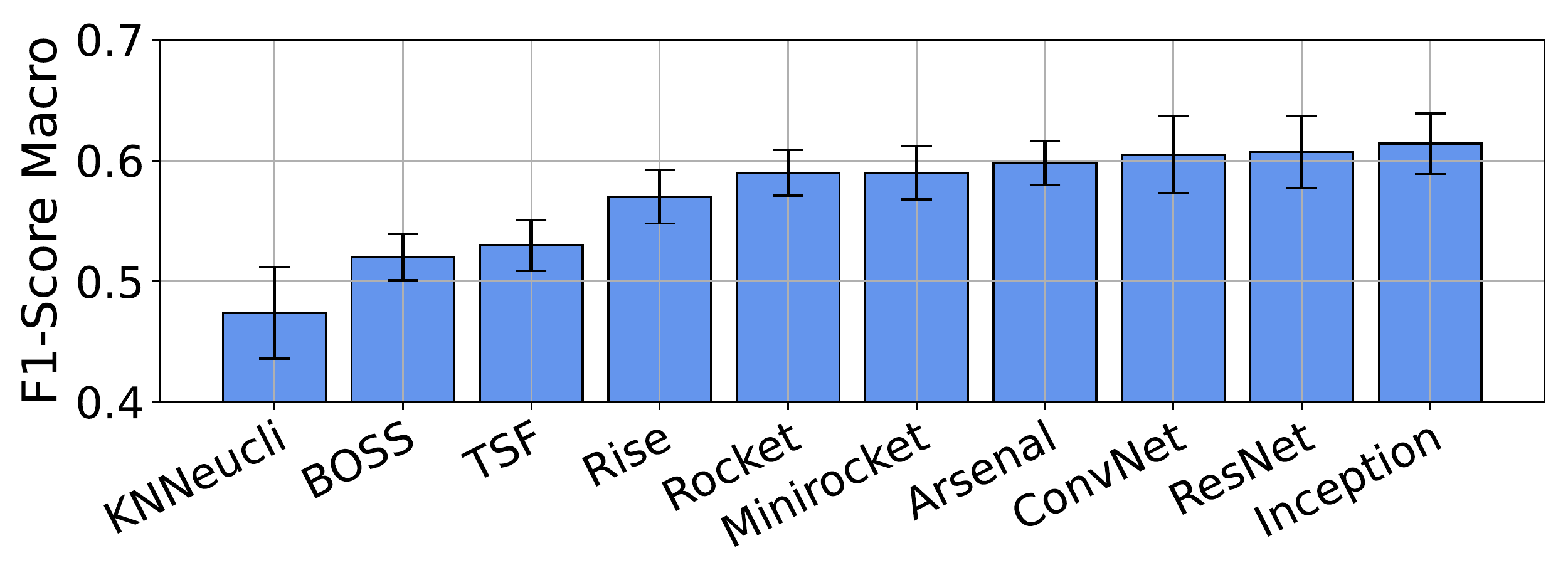}
    \vspace{-0.8cm} 
    \caption{ Average classifiers detection score through all the detection cases and all the datasets. 
    }
    \label{fig:ClfScore30min}
\end{figure}

\begin{figure}
    \centering
    \includegraphics[width=1\linewidth]{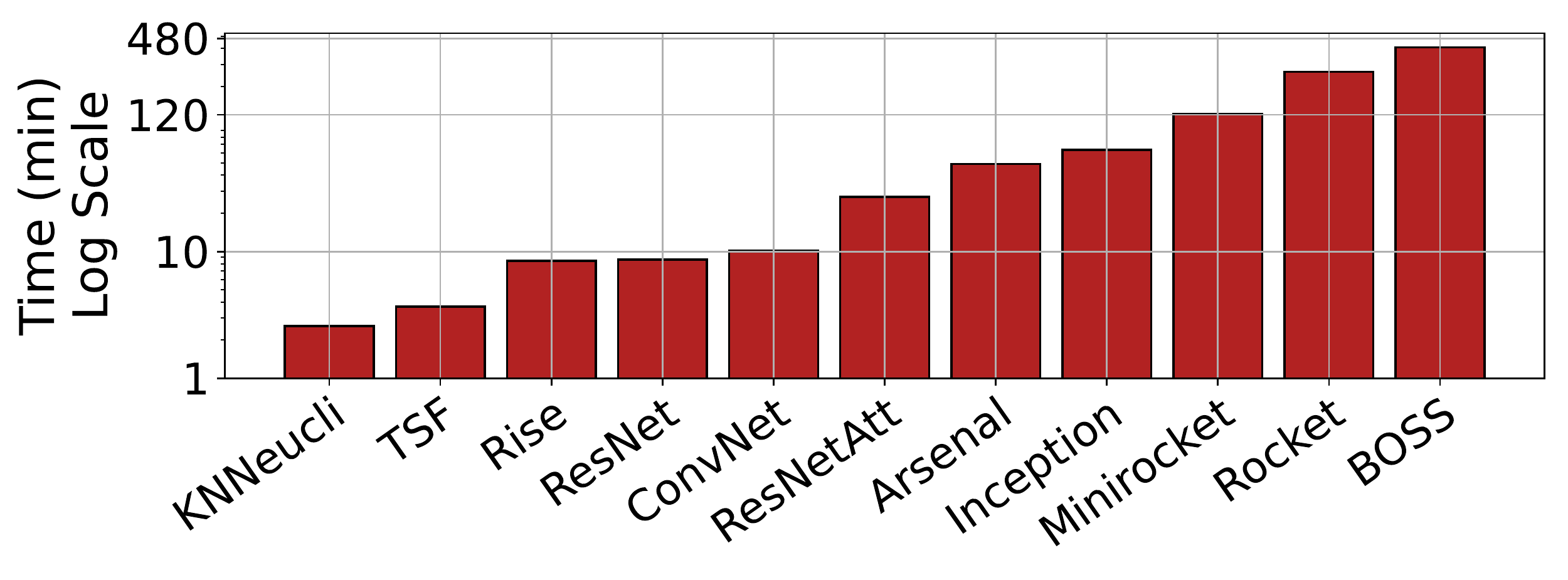}
    \vspace{-0.8cm}
    \caption{ Average running time per run (training + inference time) for all classifiers (log scale y-axis).}
    \label{fig:RunningTime}
\end{figure}

\subsection{Accuracy for 30min Sampling Frequency}
\label{subsec:halfhourresults}

The appliance detection results of the classifiers for the sampling frequency of 30min are summarized in Table~\ref{table:results30min}.  
We observe that all classifiers return poor results for the UKDALE dataset. 
(We discuss and explain these results in detail in Section~\ref{sec:dataimputation}.)
Furthermore, we note that independently of the dataset, some appliances are easier to detect than others. 
The following sections provide an analysis of these results according to the type of appliances.


\subsubsection{Tech Appliances} 
\textbf{Desktop Computer} and \textbf{Television} seem to be well detected in the REFIT dataset, with a Macro F1-Score above 0.7 for the best classifiers. 
The score obtained on \textbf{Desktop Computer} on other datasets is not as good, but is consistent with the number of time series provided. 
It can be explained by the fact that the pattern is hidden behind other appliance activation signatures in longer smart meters load curves, and thus, is hard for classifiers to detect.

\subsubsection{Kitchen Appliances} 
First, detecting \textbf{Kettle} usage looks pretty challenging, with poor results obtained by all classifiers and a Macro F1-Score $\simeq 0.45$. 
Given that a kettle operates for relatively short periods, it is understandable that its activation may not be captured using 30min sampled data. 
\textbf{Microwave} oven and classic \textbf{Oven} are not well detected in the EDF datasets. 
However, the detection score obtained on REFIT by the best two classifiers is above $0.7$, thanks to the larger amount of data available for this case in REFIT. 
Finally, the Cooker is well detected on the CER dataset.

\subsubsection{Washer Appliances} Classifiers achieve promising results detecting \textbf{Dishwasher} and \textbf{Tumble Dryer} through CER and EDF 2 datasets. 
The lower performance obtained with the EDF 1 datast is explained by the lower amount of labeled instances given for these cases. 
However, the low score results obtained on the three washer appliances for REFIT are not due to the amount of time series data. 
We believe that this poor detection score can be explained by the fact that these three devices are used in combination and have similar activation patterns; therefore, the classifiers cannot easily distinguish among them.

\subsubsection{Heating Appliances} 
The best detection scores are achieved for \textbf{Water Heater} on the EDF 1 and EDF 2 datasets. 
In France, water heaters refer mainly to devices that heat water from a hot tank, and usually operate during hours with high consumption power levels~\cite{HOHNE20191}. 
The classifiers can effectively discern this type of pattern, even using 30min sampled data. 
The lower performance on the CER dataset can be attributed to the use of two types of water heaters in Ireland: instantaneous and tank-pumped. 
Instantaneous water heaters only operate on demand, resulting in high spikes of short duration. 
Using the same label for these two devices, which have different activation signatures, significantly impacts the performance of the classifiers.
The results on heater detection are satisfying for the EDF datasets, and we assume that the score difference between EDF 1 and EDF 2 is mainly due to the span of the time period used for training the model. 
By providing a full year of electricity consumption, the model can more easily detect the heater pattern, since it trains with data during the high consumption levels of the winter season. 
The poor performance on the CER dataset for heater detection can be attributed to the fact that the heater label indicates the presence of a convector electric heater, which is typically used as a supplementary heat source in winter, rather than being the primary heat source for the home.

\subsubsection{Other Appliances} \textbf{Electric Vehicles} are well detected on the EDF 1 dataset considering the restricted number of labeled instances that we have available. 
The lengthy recharging times of electric vehicles and the high power required, combined with the fact that recharging often occurs mainly during low-consumption night-time hours, can explain the good performance we observe.

\begin{figure*}
    \includegraphics[width=0.85\linewidth]{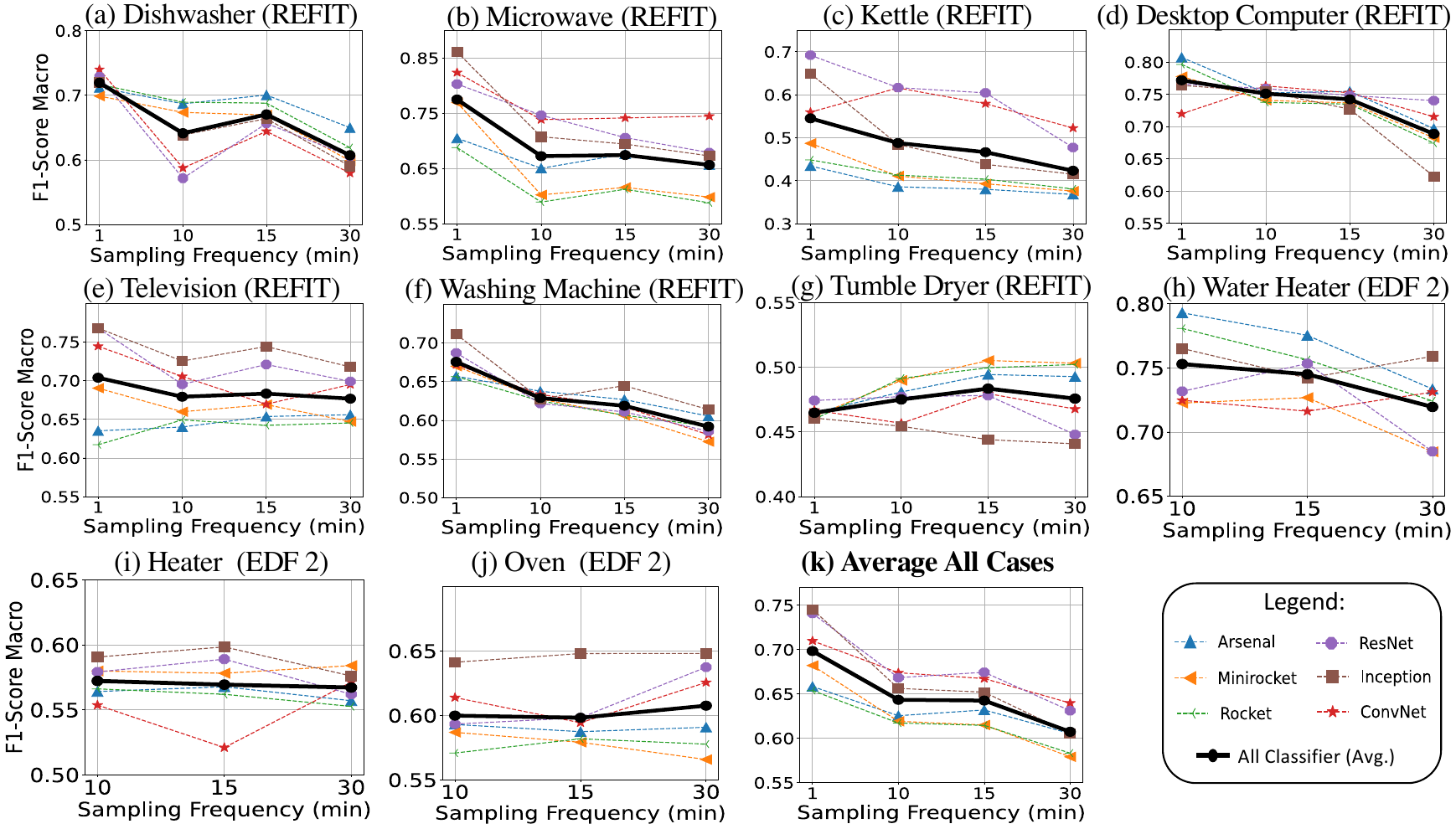}
    \caption{Influence of sampling frequency on different appliance detection cases. The detection score is given for each classifier and detection case following the resampling frequency of the data. The black line shows the average score of all the classifiers.}
    \label{fig:samplingrateimpact}
\end{figure*}

\subsubsection{Overall Classifier Results Using 30min data} 
The overall results, shown in Table~\ref{table:results30min} and Figure~\ref{fig:ClfScore30min}, demonstrate that InceptionTime outperforms other classifiers when considering the average score and rank; InceptionTime is followed by ResNet, Arsenal, ConvNet, MiniRocket and Rocket. 
Since the ResNet model enhanced with the attention mechanism was not evaluated in all the cases, we do not include it in the total average score shown in Figure~\ref{fig:ClfScore30min}. 
However, this classifier achieves relatively poor performance compared to the others (refer to Table~\ref{table:results30min}).
In light of these results, it is essential to note the difference in performance between the best and worst performing classifiers: 
convolutional-based classifiers, i.e., InceptionTime, ConvNet and ResNet, are the optimal choice for many detection cases.

Figure~\ref{fig:RunningTime} summarizes the average total running time (i.e., training and inference time together) for the 11 classifiers we studied. 
Taking into consideration the performance of the convolutional-based approaches (deep- and non deep-learning approaches), as well as their running time, we observe that this type of classifier is the most suitable for appliance detection using 30min sampled smart meter data. 
InceptionTime reaches a sligthly higher detection score, but at the cost of longer execution times. 
A balance between performance and efficiency is achieved by the ResNet and ConvNet classifiers.

\subsection{Influence of Sampling Rate}
\label{subsec:influencesamplingrate}

In this part of the experimental evaluation, we analyze the improvement of the detection score of the different classifiers, as the smart meter sampling rate increases. 
We used the REFIT and EDF 2 datasets to perform these experiments, since these datasets provide data at a higher frequency than every 30min.

Using the REFIT dataset, we performed experiments at four different sampling rates: 1min, 10min, 15min, and 30min. 
To obtain complementary results on bigger appliances that were not available with REFIT data, we also included appliance detection cases from the EDF 2 dataset. 
However, since this dataset offers data sampled at 10 min, we could only produce results for sampling rates: 10min, 15min, and 30min.

All the results are summarized in Figure~\ref{fig:samplingrateimpact}. 
For clarity, 
we only illustrate the scores of the five best classifiers. 

On average across all cases, the appliance detection accuracy decreases significantly (by almost 0.1) when the sampling rate drops from 1min to 30min.
For the best classifier (InceptionTime), the average drop is 0.15.

However, it is interesting to note that not all appliances are significantly better detected using a higher sampling frequency.
As expected, appliances that operate only for short periods, i.e., \textbf{Microwave} or \textbf{Kettle}, benefit the most when using higher smart meter frequencies. 
For example, the results in Figure~\ref{fig:samplingrateimpact} show that using 1min sampled data can significantly improve the \textbf{Kettle} detection. 
In this case, the best classifier, ResNet, achieves a 0.2 improvement in the detection score when the sampling rate increases from 30min to 1min. 
For the \textbf{Microwave} case, it is a 0.1 average gain score for all the classifiers using 1min sampled data.

Other appliances, such as \textbf{Dishwasher}, \textbf{Desktop Computer}, \textbf{Television}, \textbf{Washing Machine} and \textbf{Water Heater}, which typically operate for long periods, are better detected using higher sampling rates, as well. 
For example, using 1min level data, the \textbf{Washing Machine} is much more accurately detected than when using 30min data (refer to Figure~\ref{fig:samplingrateimpact}(f)).


\begin{figure*}
    \includegraphics[width=1\linewidth]{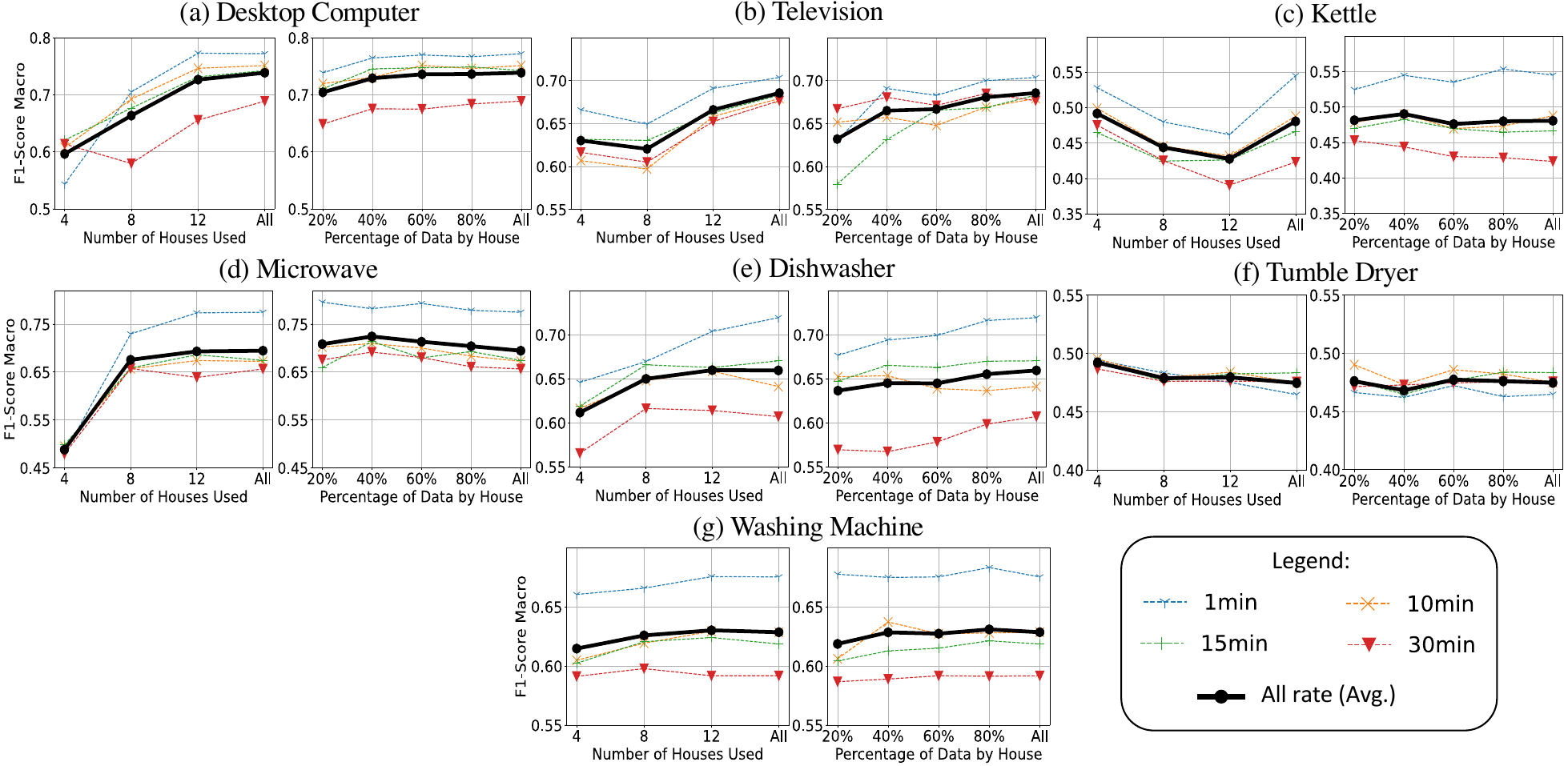}
    \vspace{-0.6cm}
    \caption{Results of the data imputation study using the REFIT dataset. For each appliance case, we can see on the left figure the evolution of the classification score according to the number of houses used, i.e., sources; on the right figure, we can see the evolution of the classification score according to the percentage of data used by houses.}
    \label{fig:DataImputationREFIT}
\end{figure*}

\subsection{Influence of Data Size}
\label{sec:dataimputation}

In this last part, 
we analyze the impact of the number of distinct households on classifier performance.
These experiments demonstrate that classifiers cannot effectively learn the patterns of an appliance using only a small number of households when the smart meter data sampling frequency is very low (this explains the poor results presented in Section~\ref{subsec:halfhourresults} for the UK-DALE dataset). 
Furthermore, we demonstrate that the number of households is more important for training the machine learning models than the amount of data available for each household. 

We compared the following two approaches for training: (i) randomly select a \emph{subset} of the houses and use all the data from these houses to train the models, and (ii) select \emph{all} houses and use a random subset of the time series from each house. 
We performed the experiments on the appliance detection cases using the REFIT dataset.
Furthermore, in order to account for the impact of the smart meter reading on these results, we performed the experiments using 4 different sampling frequencies: 1min, 10min, 15min, and 30min. 
Figure~\ref{fig:DataImputationREFIT} summarizes the results of these experiments: 
the graphs show the average performance of all classifiers\footnote{We average the performance of all classifiers listed in Table~\ref{table:results30min}, except for ResNetAtt, which could not be used with the small length of the REFIT time series.} for each sampling rate.
The black line represents the score value averaged across all sampling rates.

We note that for every sampling rate and detection case, it is almost always preferable to use all the available households and a subset of their time series, rather than to use all time series from a subset of the households. 
Indeed, data from the same house is frequently characterized by the consumption patterns of the residents. 
Instead, using data from multiple households, enables the classifier to focus on and learn the actual activation patterns of the appliances.
Interestingly, using a subset of the households, or a subset of the time series does not seem to significantly affect the detection accuracy for the \textbf{Washing Machine} and the \textbf{Tumble Dryer}.
The \textbf{Tumble Dryer} is indeed not well detected in our experiments. 
However, the detection score of the \textbf{Washing Machine} seems to be more impacted by the sampling frequency rather than by the data size.

\section{Discussion}

\begin{figure}
    \centering
    \includegraphics[width=1\linewidth]{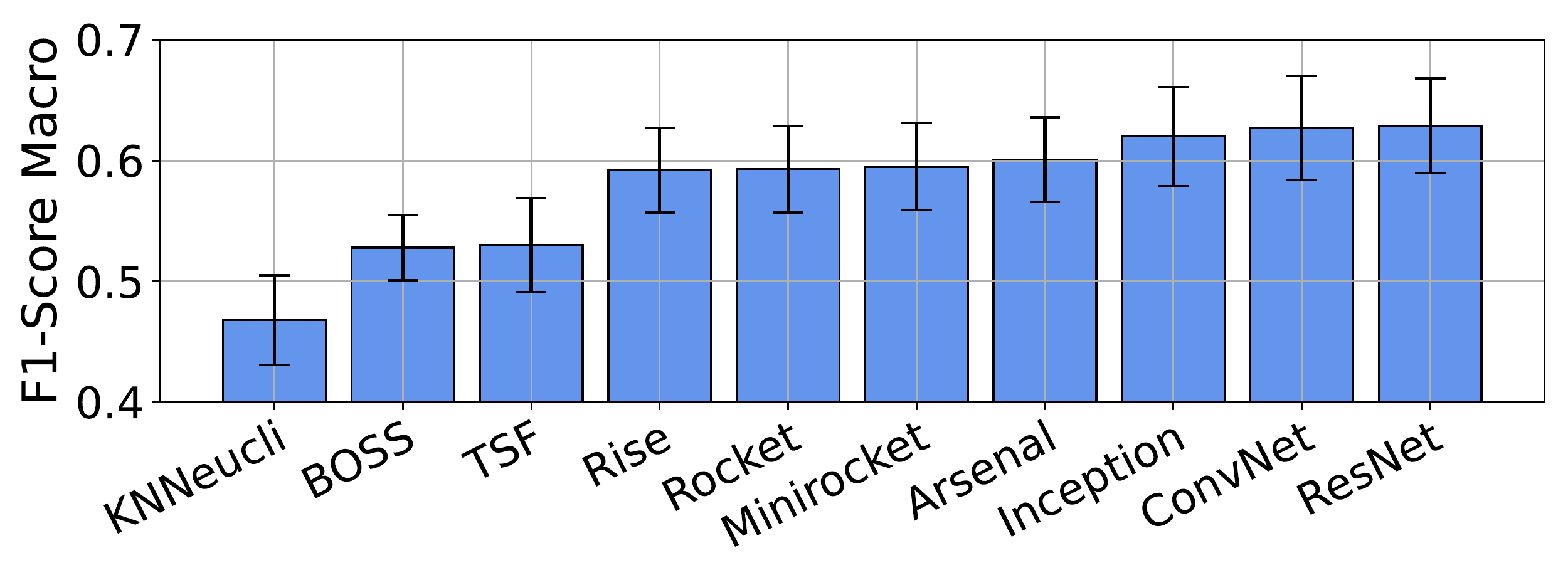}
    \vspace{-0.8cm}
    \caption{Average classifier detection score through all the experiments realized in this study (including sampling frequency influence and data size influence experiments).}
    \label{fig:ClfScoreall} 
\end{figure}
\begin{figure}
    \centering
    \includegraphics[width=1\linewidth]{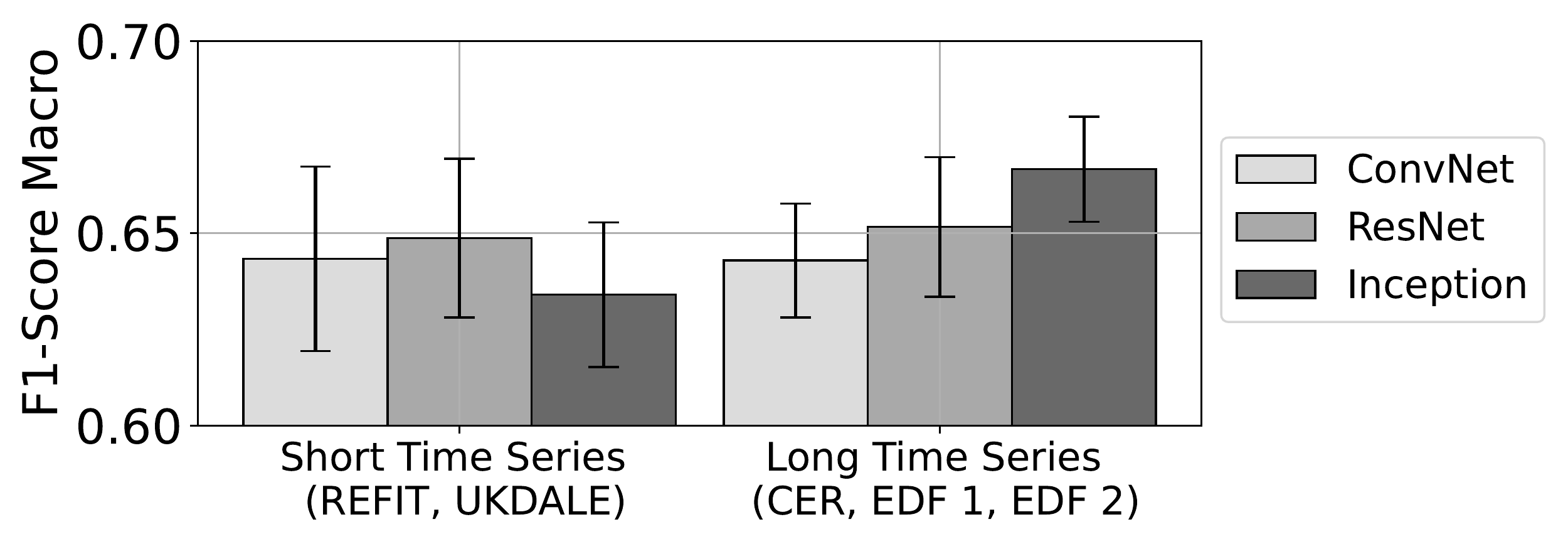}
    \vspace{-0.8cm}
    \caption{Average score of the 3 best classifiers (ConvNet, ResNet and InceptionTime) according to the time series length (i.e., datasets).}
    \label{fig:TSLenImpact}
\end{figure}

We now summarize the results of our evaluation. 
Figure~\ref{fig:ClfScoreall} shows the average score for each classifier across all the experiments conducted in our study.
The results show that the three deep learning-based methods are the most accurate overall. 
Among them, ResNet and ConvNet perform on average sligthly better than InceptionTime.
However, as shown in Figure~\ref{fig:TSLenImpact}, the average score depends on the time series length.
ResNet and ConvNet are better on average when using the short time series (REFIT and UKDALE datasets). 
InceptionTime is better on average when using long time series (CER and EDF datasets), because of InceptionTime's ability to capture long-lasting patterns through the use of a combination of differently-sized kernels.
Nevertheless, as the confidence intervals indicate, there is no clear winner among the three deep-learning classifiers. 
Based on these findings, we recommend using either ResNet or ConvNet, since their time performance is one order of magnitude faster than InceptionTime (see Figure~\ref{fig:RunningTime}).


Overall, the experiments show that for improving appliance detection, it is beneficial for electricity suppliers to collect data over extended periods of time, and at a finer time step than 30min.
Indeed, a 15min step seems to be the minimum target in order to correctly detect a certain number of appliances.
Furthermore, this study shows that further work is needed to more accurately detect appliances, even for data with 1min sampling frequency.
Nevertheless, the lack of large electricity consumption public datasets that can be used to develop and train new algorithms is an important shortcoming.
Having more good-quality data over long time intervals are necessary in order to allow for the development of more robust methods and further advancements in the field.



\section{Conclusions}

This paper presents a comprehensive evaluation of state-of-the-art time series classifiers applied to the appliance detection in very low-frequency smart meter data. 
We develop the first benchmark of time series classifiers for appliance detection using five different real datasets of very low-frequency electricity consumption with varying time series lengths. 
The results indicate that the performance of current time series classifiers varies significantly; only appliances that operate during long periods of time can be accurately detected using 30min sampled data.
However, using 1min sampling data can drastically increase the detection accuracy of small appliances.
Furthermore, deep learning-based classifiers have shown promising results in terms of accuracy, particularly for certain appliances. 
Overall, this study provides a valuable contribution to electricity suppliers, as well as analysts and practitioners, in order to help them choose the 
appropriate classifier for accurately detecting appliances in very low-frequency smart meter data.


\bibliographystyle{ACM-Reference-Format}
\bibliography{bibliography}

\end{document}